\documentstyle[aps,prb,preprint,graphicx]{revtex}

\begin{document}
\tightenlines

\title{Electron and hole states in quantum-dot quantum wells\\
within a spherical 8-band model}
\author{E. P. Pokatilov\cite{e-mail1} and V. A. Fonoberov\cite{e-mail2}}
\address{Laboratory of Multilayer Structure Physics, Department of\\
Theoretical Physics,\\
State University of Moldova, A. Mateevici\\
60, MD-2009 Chi{\c s}in{\u a}u, Moldova}
\author{V. M. Fomin\cite{perm.address} and J. T. Devreese\cite{d.address}}
\address{Theoretische Fysica van de Vaste Stof, Departement\\
Natuurkunde,\\
Universiteit Antwerpen (UIA), Universiteitsplein 1,\\
B-2610 Antwerpen, Belgi\"e}
\date{\today}
\maketitle

\begin{abstract}
In order to study heterostructures composed both of materials with strongly
different parameters and of materials with narrow band gaps, we have
developed an approach [E. P. Pokatilov {\it et al.}, Phys. Rev. B (the
preceding article in the present issue)], which combines the spherical
8-band effective-mass Hamiltonian and the Burt's envelope function
representation. Using this method, electron and hole states are calculated
in CdS/HgS/CdS/H$_{2}$O and CdTe/HgTe/CdTe/H$_{2}$O quantum-dot quantum-well
heterostructures. Radial components of the wave functions of the lowest $S$
and $P$ electron and hole states in typical quantum-dot quantum wells
(QDQWs) are presented as a function of radius. The 6-band-hole components of
the radial wave functions of an electron in the 8-band model have amplitudes
comparable with the amplitude of the corresponding 2-band-electron
component. This is a consequence of the coupling between the conduction and
valence bands, which gives a strong nonparabolicity of the conduction band.
At the same time, the 2-band-electron component of the radial wave functions
of a hole in the 8-band model is small compared with the amplitudes of the
corresponding 6-band-hole components. It is shown that in the CdS/HgS/CdS/H$%
_{2}$O QDQW holes in the lowest states are strongly localized in the well
region (HgS). On the contrary, electrons in this QDQW and both electron and
holes in the CdTe/HgTe/CdTe/H$_{2}$O QDQW are distributed through the entire
dot. The importance of the developed theory for QDQWs is proven by the fact
that in contrast to our rigorous 8-band model, there appear spurious states
within the commonly used symmetrized 8-band model.
\end{abstract}


\section{Introduction}

\label{sec:1}

Using colloidal growth technique it is possible to obtain both alloys and
multilayer quantum-dot structures. The dots are formed as free particles in
a liquid medium. In principle, they may be concentrated, and redispersed in
other hosts, such as organic polymers, to give highly doped materials.
Recently, CdS/HgS/CdS\cite{Mews1,Mews2} and CdTe/HgTe/CdTe\cite{Kershaw}
particles, termed quantum-dot quantum wells (QDQW), have been formed in
water. There are three preparation stages in the synthesis of these QDQWs:
(i) formation of the CdS (CdTe) core; (ii) substitution of the mercury ions
for the surface cadmium ions which results in a monolayer of the HgS (HgTe)
covering the core; (iii) growing the CdS (CdTe) cap layer onto the surface
of the dot. A schematic picture of typical CdS/HgS/CdS/H$_2$O and
CdTe/HgTe/CdTe/H$_2$O QDQWs and their band structure are shown in Fig.~\ref%
{fig:1}. High-resolution transmission electron microscopy images revealed%
\cite{Mews2,Kershaw} that both QDQWs are not spherical, but are
preferentially truncated tetrahedric particles. However, spherical shell
particles are commonly considered to interpret experimental data (see Refs.~%
\onlinecite{Schooss,Tkach,Chang,Jaskolski,Pokatilov1}).

The first theoretical study of QDQW electron and hole energy spectra was
made by using the parabolic approximation for the conduction and valence
bands.\cite{Schooss,Tkach,Chang} However, the position of exited hole levels
and the hole localization could not be explained in that model. Therefore,
it was necessary to take into account all six hole bands\cite{Grigoryan}.
Jaskolski and Bryant have applied\cite{Jaskolski} this model to a hole in
QDQWs. To obtain the electron spectrum, they have used the perturbation
theory for the one-band effective mass Schr\"odinger equation with an
energy-dependent mass correction. The present authors have also used\cite%
{Pokatilov1} the six-band model for a hole and the parabolic band
approximation for an electron to investigate electron, hole, impurity, and
exciton states in QDQWs. In Refs.~\onlinecite{Jaskolski,Pokatilov1} the
``symmetrized'' 6$\times$6 hole Hamiltonian has been used for the
quantum-dot heterostructure. In general, such Hamiltonian can be inadequate
for small nanocrystals (see Refs.~\onlinecite{Foreman1,Meney,Franceschi}).

The 8-band structure of bulk CdS and HgS is shown in Fig.~\ref{fig:2}, and
the 8-band structure of bulk CdTe and HgTe is shown in Fig.~\ref{fig:3}. It
is seen from these figures, that the well materials HgS and HgTe are
semimetals, i.e. they have zero band gap and the conduction band with $%
\Gamma_8$ symmetry. As also seen from Figs.~\ref{fig:2}, \ref{fig:3}, the
conduction band is parabolic only at the bottom, and even in the region of
the electron ground state energy the deviation from parabolicity is very
strong. Because of the inverted band structure, the coupling between the
conduction and valence bands should be included explicitly. In order to
simultaneously describe the nonparabolicity of the conduction band and the
influence of the conduction band on the valence bands, we consider the
8-band Pidgeon and Brown model\cite{Pidgeon} which takes into account both
the coupling of conduction and valence bands and the complex structure of
the valence band.

The exact electron and hole spectra are a starting point to calculate the
exciton spectrum and oscillator strengths of the exciton transitions. They
are required to analyze zero-phonon, one-phonon, and multi-phonon lines in
the photoluminescence spectra of these QDQWs, and to compare the obtained
results with the experiment\cite{Mews2,Kershaw,Koberling,Yeh}. In the
nonadiabatic theory\cite{Fomin} of the phonon-assisted optical transitions
in semiconductor quantum dots (QDs) the heights of the phonon peaks in the
photoluminescence spectrum strongly depend on the peculiarities of the
exciton spectrum. It has been proved for InSb\cite{Efros}, InAs\cite{Banin},
and InP\cite{Sercel} QDs that the 8-band model is important to describe the
electron and hole spectra even in spherical nanocrystals with a direct band
structure. All these facts together as well as the aforementioned inadequacy
of the previously used models make it desirable and useful to calculate the
exact electron and hole spectra using the 8-band model.

In our previous work\cite{Pokatilov2}, using the Burt's envelope function
representation\cite{Burt1,Burt3}, we have extended the 8$\times$8 model to
include heterostructures. In present paper we apply our rigorous 8-band
model for spherical quantum-dot heterostructures, namely for QDQWs. In the
next section we analyze the electron and hole states in typical QDQWs.
Conclusions are given in the Section~\ref{sec:3}.

\section{Electron and hole states for typical QDQWs}

\label{sec:2}

We consider two spherical QDQWs as shown in Fig.~\ref{fig:1}. In the
CdS/HgS/CdS/H$_2$O QDQW, the 4.4~nm in diameter CdS core is successively
covered by 1 ML of HgS (0.3~nm) and 3 MLs of CdS (0.9~nm). In the
CdTe/HgTe/CdTe/H$_2$O QDQW, the 1.5~nm in diameter CdTe core is successively
covered by 1 ML of HgTe (0.4~nm) and 3 MLs of CdTe (1.2~nm). The 8-band
energy structures of CdS/HgS and CdTe/HgTe, calculated from the bulk
effective-mass parameters listed in Table~\ref{tab}, are depicted in Figs.~%
\ref{fig:2} and \ref{fig:3} correspondingly.

To investigate electron and hole states in QDQWs we use the 8-band
effective-mass model extended to include heterostructures in Ref.~%
\onlinecite{Pokatilov2}. In spherical quantum-dot heterostructures, namely
in spherical QDQWs, all effective-mass parameters entering the 8-band
Hamiltonian depend only on the radial coordinate $r$. Therefore, electron
and hole states are eigenfunctions of the total angular momentum $j$ and its 
$z$-component $m\equiv j_z$. Consequently, the electron or hole wave
function can be written as a linear expansion in the eight Bloch functions $%
u_{J,\mu}^{c(v)}$ ($u_{J,\mu}^c$ and $u_{J,\mu}^v$ are the Bloch functions
of the conduction and valence bands from Refs.~\onlinecite{Efros,Pokatilov2}%
, $J$ is the Bloch function angular momentum, and $\mu\equiv J_z$ is its $z$%
-component) as 
\begin{equation}  \label{wf}
\Psi_{j,m}({\bf r})=\sum\limits_{\mu=-1/2}^{1/2} F^{c;j,m}_{1/2,\mu}({\bf r}%
)\,u_{1/2,\mu}^c + \sum\limits_{J=1/2}^{3/2}\sum\limits_{\mu=-J}^{J}
F^{v;j,m}_{J,\mu}({\bf r})\,u_{J,\mu}^v,
\end{equation}
where the envelope functions $F^{c(v);j,m}_{J,\mu}({\bf r})$ are 
\begin{eqnarray}  \label{envelopes}
F^{c;j,m}_{1/2,\mu}({\bf r})&=&\sum\limits_{l=j-1/2}^{j+1/2}\sum\limits_{%
\lambda=-l}^{l} C_{1/2,\mu;l,\lambda}^{j,m}\,R^{c;j}_{1/2,l}(r)\,
Y_{l,\lambda}(\theta,\phi),  \nonumber \\
F^{v;j,m}_{3/2,\mu}({\bf r})&=&i^{\mu-3/2}\sum\limits_{l=|j-3/2|}^{j+3/2}%
\sum\limits_{\lambda=-l}^{l}
C_{3/2,\mu;l,\lambda}^{j,m}\,R^{v;j}_{3/2,l}(r)\, Y_{l,\lambda}(\theta,\phi),
\\
F^{v;j,m}_{1/2,\mu}({\bf r})&=&i^{1/2-\mu}\sum\limits_{l=j-1/2}^{j+1/2}\sum%
\limits_{\lambda=-l}^{l} C_{1/2,\mu;l,\lambda}^{j,m}\,R^{v;j}_{1/2,l}(r)\,
Y_{l,\lambda}(\theta,\phi).  \nonumber
\end{eqnarray}
Here, $R^{c(v);j}_{J,l}(r)$ are the radial envelope functions, $%
C_{J,\mu;l,\lambda}^{j,m}$ are the Clebsch-Gordan coefficients, and $%
Y_{l,\lambda}(\theta,\phi)$ are the spherical harmonics. In Eqs.~(\ref{wf}),
(\ref{envelopes}), $F^{c;j,m}_{1/2,\mu}({\bf r})$ are the 2-band-electron
components of the 8-band wave function $\Psi_{j,m}({\bf r})$ (referred to as
electron components in the following) and $F^{v;j,m}_{J,\mu}({\bf r})$ are
the 6-band-hole components of the 8-band wave function $\Psi_{j,m}({\bf r})$
(referred to as hole components in the following). The electron or hole
eigenenergy $E_j$, corresponding to the wave function $\Psi_{j,m}({\bf r})$,
does not depend on $m$, because within the spherical approximation the
energy spectrum is degenerate with respect to the $z$-component of the total
momentum. It is also known\cite{Pokatilov2} that the parity $p$ is conserved
in the spherical approximation.

For electron and hole levels, obtained within the spherical 8-band model, we
use a common notation: $nQ_j^{(e)}$ denotes an electron state and $nQ_j^{(h)}
$ denotes a hole state, where $n$ is the number of the level with a given
symmetry and $Q=S,P,D,\dots$ is the lowest value of the momentum $l$ in the
spherical harmonics of Eqs.~(\ref{wf}), (\ref{envelopes}) in front of the CB
Bloch functions for an electron state and in front of the VB Bloch functions
for a hole state, i.e. $Q=j-p/2$ for an electron and $Q=\min(j+p/2,|j-3p/2|)$
for a hole.

Parameters $\xi(r)$ and $\chi(r)$ introduced in Ref.~\onlinecite{Pokatilov2}
are responsible for the nonsymmetrical form of the 8-band Hamiltonian. $\chi$
is explicitly defined through the effective-mass parameters of the
homogeneous bulk model: 
\begin{equation}  \label{chi8}
\chi=(5\gamma-\gamma_1-1)/3.
\end{equation}
Following Refs.~\onlinecite{Pokatilov2,Foreman2} we take for $\xi$ 
\begin{equation}  \label{xi8}
\xi=v.
\end{equation}

Using our rigorous 8-band model and the aforementioned set of parameters,
the lowest $S$ and $P$ electron and hole states in both QDQWs are analyzed.
The $1S_{1/2}^{(e)}$, $1P_{3/2}^{(e)}$, $1P_{1/2}^{(e)}$ electron and $%
1P_{3/2}^{(h)}$, $1P_{1/2}^{(h)}$, $1S_{3/2}^{(h)}$, $2S_{3/2}^{(h)}$, $%
1S_{1/2}^{(h)}$, $1P_{5/2}^{(h)}$, $2P_{5/2}^{(h)}$, $2P_{3/2}^{(h)}$, $%
3S_{3/2}^{(h)}$ hole energy levels are presented and the corresponding
radial wave functions are plotted as a function of $r$ in Figs.~\ref{fig:4}
and \ref{fig:5}, respectively, for CdS/HgS/CdS/H$_2$O and CdTe/HgTe/CdTe/H$_2
$O QDQWs. In these figures, the following denotations\cite{Pokatilov2} 
\begin{equation}  \label{otherR}
\begin{array}{rclrcl}
R^{c;j}_{1/2,j-1/2} & = & R^+_{c,j}, & R^{c;j}_{1/2,j+1/2} & = & -R^-_{c,j},
\\ 
R^{v;j}_{3/2,j+1/2} & = & R^+_{h1,j}, & R^{v;j}_{3/2,j-1/2} & = & R^-_{h1,j},
\\ 
R^{v;j}_{3/2,j-3/2} & = & -R^+_{h2,j}, & R^{v;j}_{3/2,j+3/2} & = & 
-R^-_{h2,j}, \\ 
R^{v;j}_{1/2,j+1/2} & = & R^+_{s,j}, & R^{v;j}_{1/2,j-1/2} & = & R^-_{s,j}.%
\end{array}%
\end{equation}
are used for the radial components $R_{c}(r)$ (solid curves), $R_{h1}(r)$
(dashed curves), $R_{h2}(r)$ (dotted curves), and $R_{s}(r)$ (dash-dotted
curves). For all eigenstates, the contribution to the normalization integral 
$\int_0^{\infty} r^2
dr\left(R^2_{c}(r)+R^2_{h1}(r)+R^2_{h2}(r)+R^2_{s}(r)\right)=1$ from each
particular radial component is indicated in percent. The vertical lines in
Figs.~\ref{fig:4}, \ref{fig:5} represent the spherical borders between the
different materials.

The function $R_c(r)$ is called further the electron radial component of the
wave function because in Eqs.~(\ref{wf}), (\ref{envelopes}) it is written in
front of the CB Bloch functions. The functions $R_{h1}(r)$, $R_{h2}(r)$, and 
$R_s(r)$ are called further the hole radial components of the wave function
because in Eqs.~(\ref{wf}), (\ref{envelopes}) they are written in front of
the VB Bloch functions. If the coupling of the conduction and valence bands
is not considered, the electron wave function is described by the first term
in Eq.~(\ref{wf}) (i.e. by the electron radial component), and the hole wave
function is described by the second term in Eq.~(\ref{wf}) (i.e. by the hole
radial components). The inclusion of the coupling between the conduction and
valence bands (which is realized in the 8-band model) results in the fact
that the electron wave function includes the 6-band-hole components (HE), in
addition to the 2-band-electron component (EE), and the hole wave function
includes the 2-band-electron component (EH), in addition to the 6-band-hole
components (HH).

The main radial component for the electron wave functions is EE. As seen
from Figs.~\ref{fig:4}, \ref{fig:5}, the amplitude of HE is about 1/2 of EE.
It is also seen, that about 20\% of the contribution into the normalization
integral comes from HE. Such a big contribution from the hole components is
because of the strong nonparabolicity of the conduction band in the well
materials (see Figs.~\ref{fig:2}, \ref{fig:3}). The spin-orbit splitting of
the electron levels, when going from the one-band model to the 8-band model,
is very small. For example, the level $1P^{(e)}$ splits into levels $%
1P_{1/2}^{(e)}$ and $1P_{3/2}^{(e)}$ with the distance between them only
0.8~meV for CdS/HgS/CdS/H$_2$O QDQW and 3.6~meV for CdTe/HgTe/CdTe/H$_2$O
QDQW. It should be noted, that this splitting occurs in such way, that the
lowest level in the former QDQW is $1P_{3/2}^{(e)}$ and the lowest level in
the latter QDQW is $1P_{1/2}^{(e)}$. The electron ground state energy is not
split and it is $1S_{1/2}^{(e)}$ for both QDQWs.

Three HH are the main components in the hole wave functions. The amplitude
of EH here is about 1/10 of the amplitudes of HH, and this component gives
very small contribution into the normalization integral. Comparing with the
six-band model, no additional splitting of the hole levels occurs in the
8-band model. Although the hole ground state remains $1P_{3/2}^{(h)}$ in the
CdS/HgS/CdS/H$_2$O QDQW, the position of the hole levels changes within this
model (compare with Ref.~\onlinecite{Jaskolski}). The hole ground energy in
the CdTe/HgTe/CdTe/H$_2$O QDQW is $1S_{3/2}^{(h)}$.

One can see that the number of electron levels in an arbitrary energy
interval is less than the number of hole levels in the same interval. This
is connected with the complex structure of the valence band and with the
fact that the hole effective masses are larger than the electron effective
mass. The positions of an electron and a hole differ significantly in the
CdS/HgS/CdS/H$_2$O QDQW. While the electron is distributed through the
entire dot, the hole is almost completely localized in the HgS layer. It is
worth pointing out that in this QDQW, the electron quantization energy is
several times larger than the hole quantization energy. Therefore, while the
holes do not practically penetrate into the exterior medium, the probability
of the electron presence in H$_2$O is about 5\%. In the CdTe/HgTe/CdTe/H$_2$%
O QDQW, the electron and hole behave similarly: both charge carriers are
distributed through the entire dot and the probability of their presence in H%
$_2$O is about 5\%. This difference between the considered QDQWs can be
explained by the fact that the lowest energy levels in the CdS/HgS/CdS/H$_2$%
O QDQW lie inside the HgS well, while the lowest energy levels in the
CdTe/HgTe/CdTe/H$_2$O QDQW lie outside the HgTe well (see Figs.~\ref{fig:2}, %
\ref{fig:3}).

Finally, taking the CdS/HgS/CdS/H$_{2}$O QDQW as an example, we examine,
first, what error occurs if one uses the symmetrized 8-band Hamiltonian to
find electron and hole states, and, second, how electron and hole states can
change if one supposes that HgS is not a semimetal with $E_{g}=-190$~meV,
but a narrow-gap semiconductor with $E_{g}=200$~meV.

The use of the symmetrized 8-band Hamiltonian leads to the following changes
in the electron and hole spectra as compared to the spectra obtained within
our rigorous approach. (i) For every value of total angular momentum and
parity, there appears one spurious electron level with energy about $1080$%
~meV, $1100$~meV, $1120$~meV, \dots\ for the $S$, $P$, $D$, \dots\ electron
states, respectively. These electron states are spurious, because about $99$%
\% of the electron density is concentrated in a very narrow region (less
than $0.5$~nm wide) near the boundary between the QDQW and water. (ii) At
the same time, the energy of the lowest genuine electron levels increases by 
$66$~meV, $43$~meV, $11$~meV, \dots\ for the $S$, $P$, $D$, \dots\ electron
states, respectively. Since all other electron states must be orthogonal to
the lowest --- spurious \ --- state, a considerable part of the low-lying
electron states possess a high probability (of about $1/2)$ that the
electron is in water. (iii) The energy of hole states changes only slightly
(by $\pm 5$~meV). The density of the lowest hole states is also slightly
modified.

The increase of the value $E_{g}$(HgS) from $-190$~meV to $200$~meV, while
all other parameters are kept unchanged, results in the following. (i) The
energy of all the lowest electron levels increases by about $80$~meV. This
fact leads to a decrease of the electron density in the HgS well and to a
corresponding increase of the electron density in the CdS layers. (ii) The
energy of hole levels changes weakly (by less than $1$~meV), and the
distribution of the hole density is practically unchanged.

\section{Conclusions}

\label{sec:3}

A spherical QDQW structure is considered for the first time within the
8-band model. This approach generalizes the one-band and the six-band
models, which have been used for QDQWs so far, and exactly takes into
account the nonparabolicity of the electron dispersion law. Electron and
hole energy spectra and the corresponding wave functions for typical QDQWs
have been examined. While in the CdS/HgS/CdS/H$_2$O QDQW (recently
considered in the framework of the 6-band model\cite{Jaskolski}) holes in
the lowest states are strongly localized in the well region, in the
CdTe/HgTe/CdTe/H$_2$O QDQW holes are shown to be distributed through the
entire dot. At the same time, electrons in both types of QDQWs are not
localized in the quantum well.

The lowest optically active electron-hole pair state in both QDQWs is $%
1S_{1/2}^{(e)}$-$1S_{3/2}^{(h)}$. The energy of this state is 2.022~eV for
the CdS/HgS/CdS/H$_2$O QDQW and 1.723~eV for the CdTe/HgTe/CdTe/H$_2$O QDQW.
The inclusion of the exciton effect reduces each of these energies by about
100~meV and makes them close to the experimental values given in Refs.~%
\onlinecite{Mews2,Kershaw} (the analysis of this effect is in progress). The
aforementioned conclusions about the properties of the electron and hole
spectra are a consequence of 1 monolayer thickness of the HgS (HgTe) well
and of the semimetal character of the well material. We have made it certain
that the symmetrized 8-band model is not capable to describe electronic
states in QDQWs composed of materials with very dissimilar band parameters.
It has been also shown that the obtained results do not depend critically on
the actual value of the well band gap. In summary, we have demonstrated that
the developed model is an effective tool to analyze quantum-dot
heterostructures, which include thin well layers and also narrow-band-gap
materials.

\subsection*{ACKNOWLEDGEMENTS}

This work has been supported by the GOA BOF UA 2000, IUAP, FWO-V projects
G.0287.95, G.0274.01N, W.O.G. WO.025.99N (Belgium), and CRDF Award MP2-2281
(Moldova).


\begin{table}[tbp]
\caption{8-band bulk effective-mass parameters of the QDQWs constituent
materials.}
\label{tab}%
\begin{tabular}{cccccc}
Parameters & HgS & CdS & HgTe & CdTe & H$_2$O \\ \hline
$E_p\,(\text{eV})$ & 13.2\tablenotemark[1] & 21.0\tablenotemark[1] & 15.5%
\tablenotemark[2] & 17.4\tablenotemark[6] & 0\tablenotemark[9] \\ 
$E_g\,(\text{eV})$ & $-$0.190\tablenotemark[1] & 2.56\tablenotemark[1] & $-$%
0.32\tablenotemark[3] & 1.57\tablenotemark[7] & 8.0\tablenotemark[9] \\ 
$\Delta\,(\text{eV})$ & 0.07\tablenotemark[1] & 0.07\tablenotemark[1] & 0.91%
\tablenotemark[3] & 0.953\tablenotemark[8] & 0\tablenotemark[9] \\ 
$E_v\,(\text{eV}) $ & 0 & $-$0.93\tablenotemark[1] & --- & --- & $-$4.0%
\tablenotemark[9] \\ 
$E_v\,(\text{eV})$ & --- & --- & 0 & $-$0.53\tablenotemark[7] & $-$4.6%
\tablenotemark[10] \\ 
$\alpha$ & $-$1.0\tablenotemark[1] & $-$2.57\tablenotemark[1] & 1.0%
\tablenotemark[4] & 1.30\tablenotemark[8] & 1\tablenotemark[9] \\ 
$\gamma_1$ & 0.35\tablenotemark[1] & $-$1.02\tablenotemark[1] & 1.66%
\tablenotemark[5] & 1.68\tablenotemark[8] & 1\tablenotemark[9] \\ 
$\gamma$ & $-$0.67\tablenotemark[1] & $-$0.75\tablenotemark[1] & $-$0.31%
\tablenotemark[5] & 0.01\tablenotemark[8] & 0\tablenotemark[9]%
\end{tabular}
\tablenotetext[1]{Ref.~\onlinecite{Pokatilov2}} \tablenotetext[2]{Ref.~%
\onlinecite{Guldner}} \tablenotetext[3]{Ref.~\onlinecite{Orlowsky}} %
\tablenotetext[4]{Ref.~\onlinecite{Truchsess}} \tablenotetext[5]{Ref.~%
\onlinecite{Gashimzade}} \tablenotetext[6]{Ref.~\onlinecite{Redigilo}} %
\tablenotetext[7]{Ref.~\onlinecite{Eich}} \tablenotetext[8]{Ref.~%
\onlinecite{Aliev}} \tablenotetext[9]{Ref.~\onlinecite{Jaskolski}} %
\tablenotetext[10]{Ref.~\onlinecite{Nethercot}}
\end{table}



\begin{figure}[tbp]
\caption{A schematic picture of typical CdS/HgS/CdS (left-hand-side panel)
and CdTe/HgTe/CdTe (right-hand side panel) QDQWs in H$_2$O. Diameters of the
considered QDQWs are 6.8~nm and 4.7~nm correspondingly. The band structure
of each QDQW is shown below.}
\label{fig:1}
\end{figure}

\begin{figure}[tbp]
\caption{The bulk 8-band structure of CdS and HgS calculated using the
material parameters from Table~\ref{tab}. Dashed horizontal lines denote the
lowest $S$ and $P$ electron and hole energy levels in a typical CdS/HgS/CdS/H%
$_2$O QDQW (diameter = 6.8~nm). The corresponding radial components of the
wave functions are presented as a function of $r$ in Fig.~\ref{fig:4}.}
\label{fig:2}
\end{figure}

\begin{figure}[tbp]
\caption{The bulk 8-band structure of CdTe and HgTe calculated using the
material parameters from Table~\ref{tab}. Dashed horizontal lines denote the
lowest $S$ and $P$ electron and hole energy levels in a typical
CdTe/HgTe/CdTe/H$_2$O QDQW (diameter = 4.7~nm). The corresponding radial
components of the wave functions are presented as a function of $r$ in Fig.~%
\ref{fig:5}.}
\label{fig:3}
\end{figure}

\begin{figure}[tbp]
\caption{The lowest $S$ and $P$ electron and hole energy levels and
corresponding radial components of the wave functions for a typical
CdS/HgS/CdS/H$_2$O QDQW. The contribution to the normalization integral from
each radial component is indicated. Vertical lines represent the spherical
boundaries which separate the CdS core, the HgS monolayer, the CdS cap
layer, and the surrounding medium.}
\label{fig:4}
\end{figure}

\begin{figure}[tbp]
\caption{The lowest $S$ and $P$ electron and hole energy levels and
corresponding radial components of the wave functions for a typical
CdTe/HgTe/CdTe/H$_2$O QDQW. The contribution to the normalization integral
from each radial component is indicated. Vertical lines represent the
spherical boundaries which separate the CdTe core, the HgTe monolayer, the
CdTe cap layer, and the surrounding medium.}
\label{fig:5}
\end{figure}


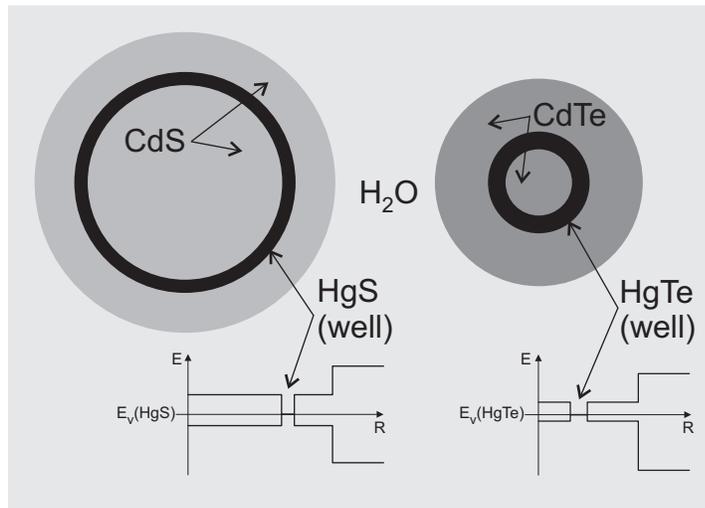
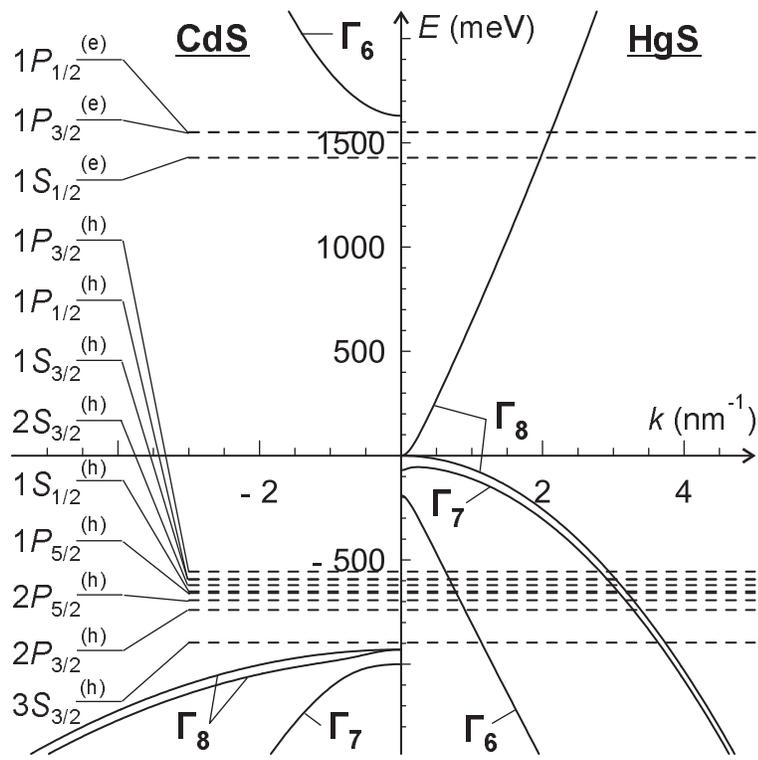
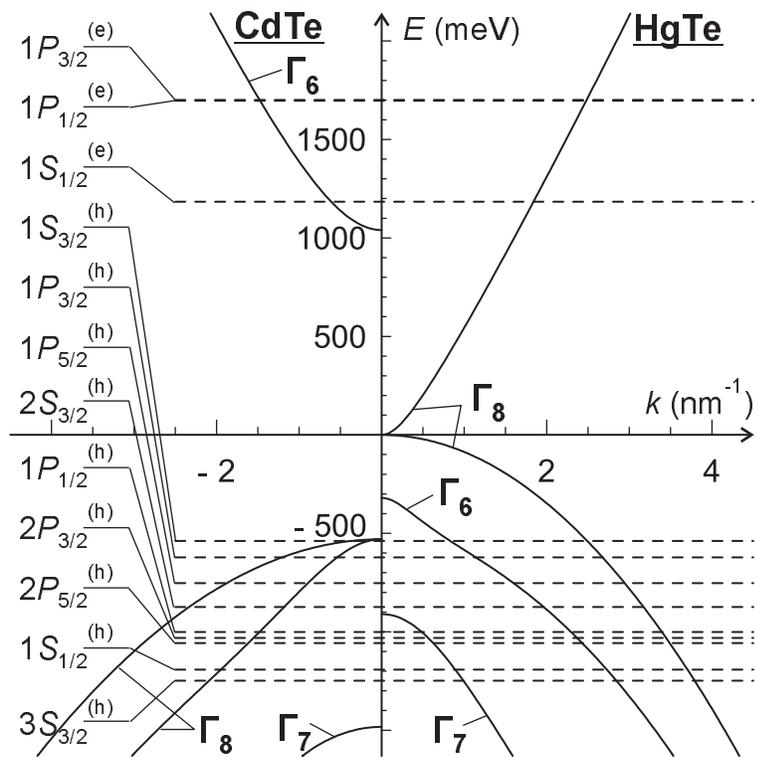
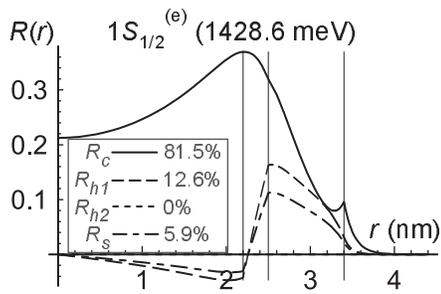
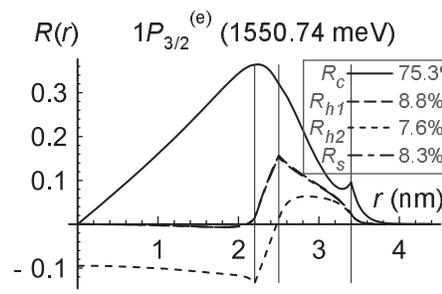
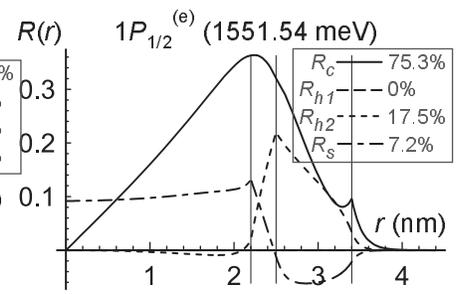
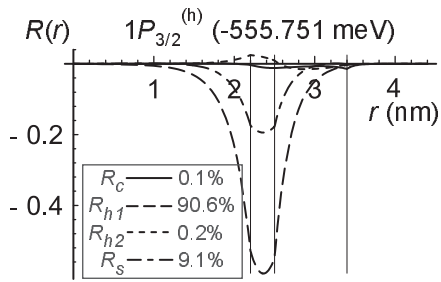
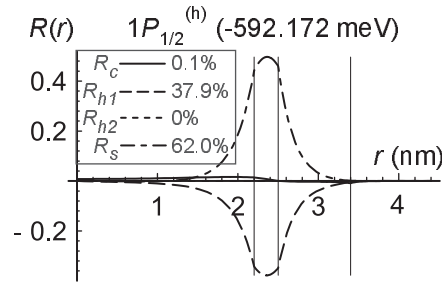
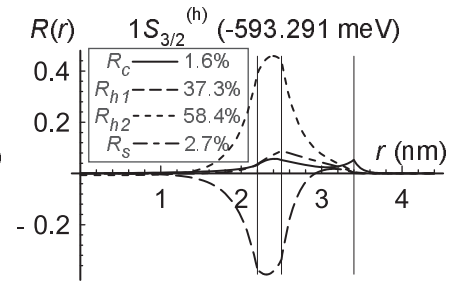
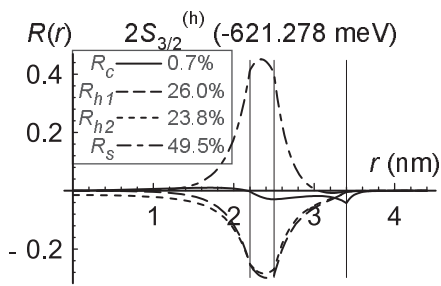
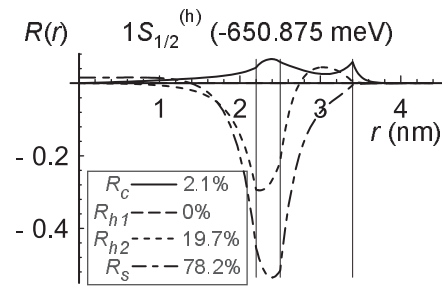
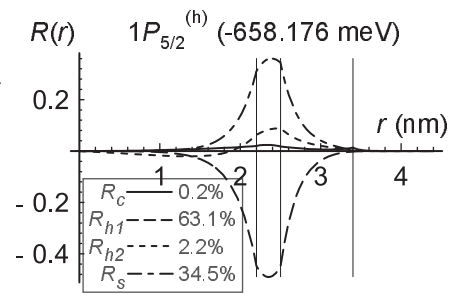
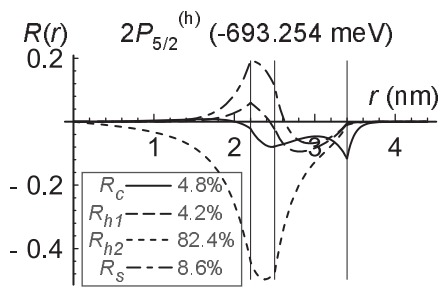
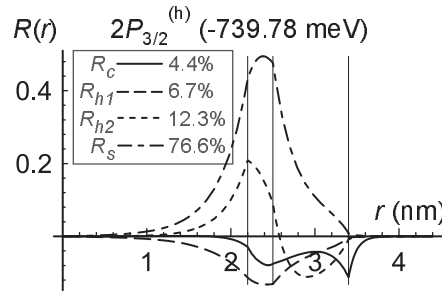
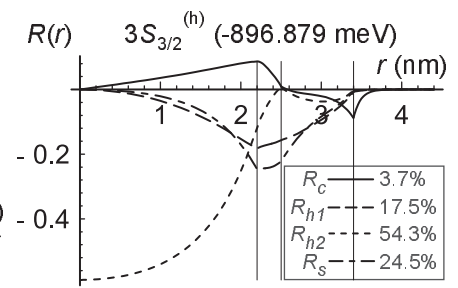
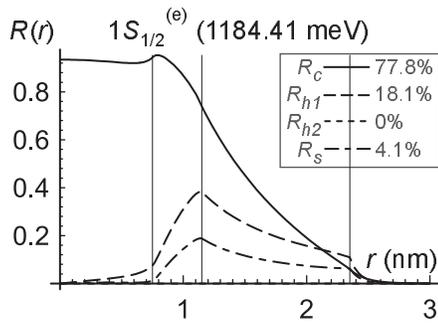
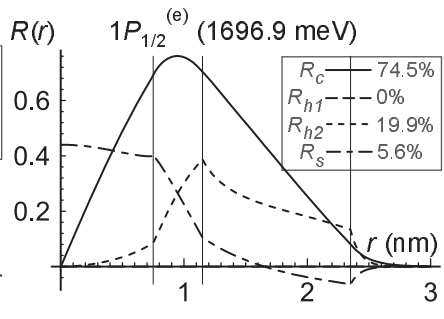
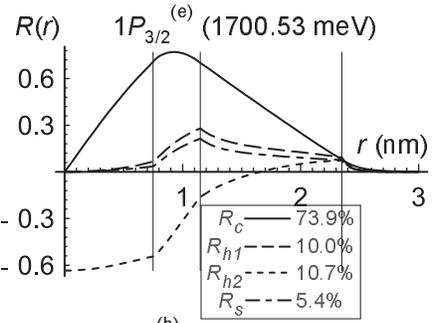
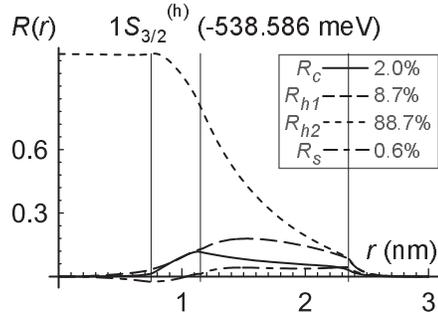
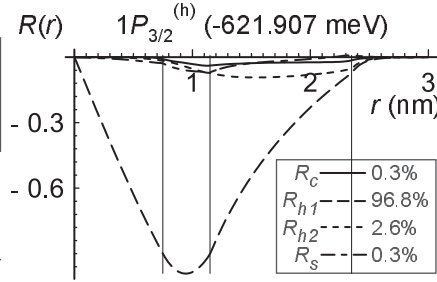
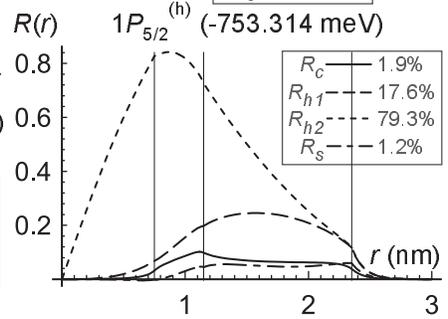
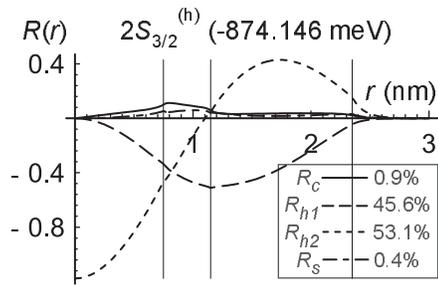
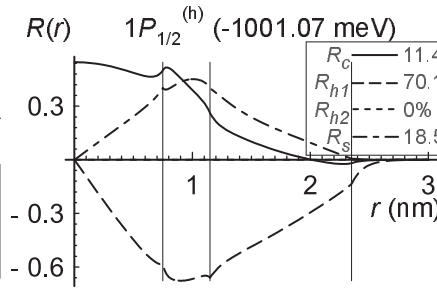
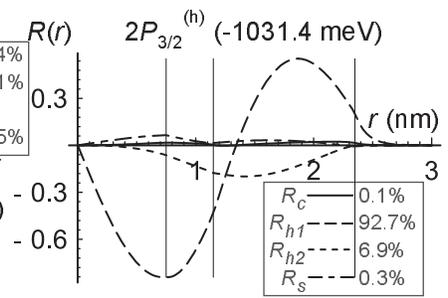
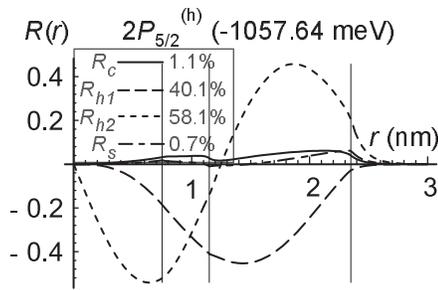
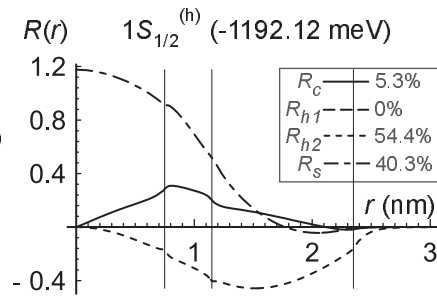
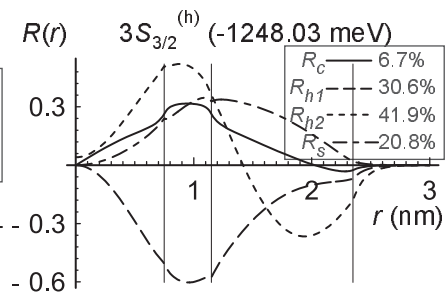
\end{document}